\documentclass[3p,times]{elsarticle}

\usepackage{ecrc}
\usepackage{graphicx}

\volume{00}

\firstpage{1}

\journalname{Nuclear Physics A}

\runauth{}

\jid{nupha}

\usepackage{amssymb}

 \biboptions{compress}

\usepackage[figuresright]{rotating}

\begin{document}

\begin{frontmatter}



\dochead{}

\title{Open heavy flavour and J/$\psi$ production in proton-proton collisions measured with the ALICE experiment at LHC}



\author{Claudio GEUNA (for the ALICE Collaboration)}
\address{Commissariat $\grave{a}$ l'Energie Atomique, Saclay, IRFU/SPhN, 91191 Gif-sur-Yvette, France}

\ead{claudio.geuna@cea.fr}

\begin{abstract}
Open heavy flavour and J/$\psi$ production are powerful tools to test pQCD predictions in proton-proton (pp) collisions in the new LHC energy regime. In addition, they provide the necessary reference for the ALICE heavy-ion program. The ALICE experiment at the LHC collected pp collision data at $\sqrt{s}$ = 7 and 2.76 TeV in 2010 and 2011. We report here the latest results of open heavy flavour and J/$\psi$ production at both mid- and forward-rapidity via several hadronic and leptonic decay channels. These include, among other topics, the first LHC result on J/$\psi$ polarization, multiplicity dependence of the J/$\psi$ production, D-meson measurement down to low transverse momentum ($p_{\rm T}$) and J/$\psi$ production from B-hadron decays. Comparisons with different theoretical models will be discussed as well. 

\end{abstract}


\end{frontmatter}


\section{Introduction}
\label{11}

The study of open heavy flavour and quarkonium production in pp collisions has a twofold interest.\par 
On one hand, these measurements are a fundamental testing ground of QCD calculations provided by several theoretical models in the new LHC energy regime ($\sqrt{s}$ = 7 and 2.76 TeV). For the open flavours, priority has been given to the measurement of the production cross section of charm and beauty hadrons (down to low $p_{\rm T}$) and to compare the experimental results with pQCD predictions. For quarkonia, the aim is to understand better the production of quarkonium states which is still a challenge for theoretical models. On the other hand, pp collisions at $\sqrt{s}$~=~2.76~TeV provide a crucial reference for the measurements performed in Pb-Pb collisions at the same $\sqrt{s}$ per nucleon pair.     
\par
ALICE~\cite{ALICEcoll} is the LHC experiment dedicated to heavy-ion studies. The setup includes a central barrel (-0.9~$<$~$\eta$ $<$~0.9) and a forward muon spectrometer (-4 $<$ $\eta$ $<$ -2.5). The barrel detectors, all located inside a large solenoidal magnet, provide momentum measurement for charged particles with $p_{\rm T}$ $>$ 100 MeV/c and particle identification up to $p_{\rm T}$ $\approx$ 10 GeV/c. The detectors, used in this analysis, are the Inner Tracking System (ITS), the Time Projection Chamber (TPC), the Transition Radiation Detector (TRD), the Time-Of-Flight (TOF) detector and the Electromagnetic Calorimeter (EMCal). The ITS~\cite{ITS-key} consists of six layers of silicon detectors surrounding the beam pipe and it is used for the vertex determination and track reconstruction close to the interaction point (IP). The TPC~\cite{TPC-key}, a large cylindrical drift detector, is the main tracking device in the central barrel and it provides also particle identification. The TPC is surrounded by the Transition Radiation Detector (TRD) which provides a good separation of electrons from pions, in particular for momenta above 1 GeV/c~\cite{TRD-key}. Finally at larger radii, at a distance of 3.7 m from the beam axis, the Time-Of-Flight (TOF) detector~\cite{TOF-key} serves to extend significantly the momentum range of particle identification and an Electromagnetic Calorimeter (EMCal)~\cite{EMCal-key} is located at 4.5 m from the beam line.\par
Muons with a momentum larger than 4 GeV/c are detected in the forward muon spectrometer. It consists of a front absorber, a dipole magnet providing a field integral of 3 Tm, and tracking and trigger chambers. Tracking is performed by means of five tracking stations, each composed of two planes of Cathode Pad Chambers. The trigger system consists of two stations, each of which consists of two planes of Resistive Plate Chambers placed behind a thick iron wall located downstream of the muon tracking system. Finally, two forward detectors (VZERO) are used for triggering inelastic pp interactions and for the rejection of beam-gas events. They consist of scintillator arrays covering the pseudorapidity ranges -3.7~$<$~$\eta$~$<$~-1.7 and 2.8~$<$~$\eta$~$<$~5.1.\par
In these proceedings, we present results on open heavy flavour and J/$\psi$ production based on 2010 (2011) pp data at $\sqrt{s}$~=~7 (2.76)~TeV. The data sample corresponds to events collected with the minimum bias (MB) pp trigger which is defined as the logical OR between two conditions: at least one hit in the ITS Silicon Pixel Detector (SPD) and a signal in at least one of the two VZERO detectors. For the muon analysis, a more restrictive trigger is used ($\mu$-MB). It requires at least one hit in the muon trigger system in addition to the MB trigger requirements.         


\section{Results on Open heavy flavour production}
\label{11}

In ALICE, open heavy flavour production can be studied in two different pseudorapidity regions. In the central barrel, the study can be performed by measuring charmed D-meson production and electrons coming from semileptonic decays of heavy flavour hadrons. On the other hand, in the forward muon spectrometer the production of muons from heavy flavour hadron decays is measured.\par
\begin{figure}[htbp]
\begin{center}
\includegraphics[width=7.5cm,height=6.33cm]{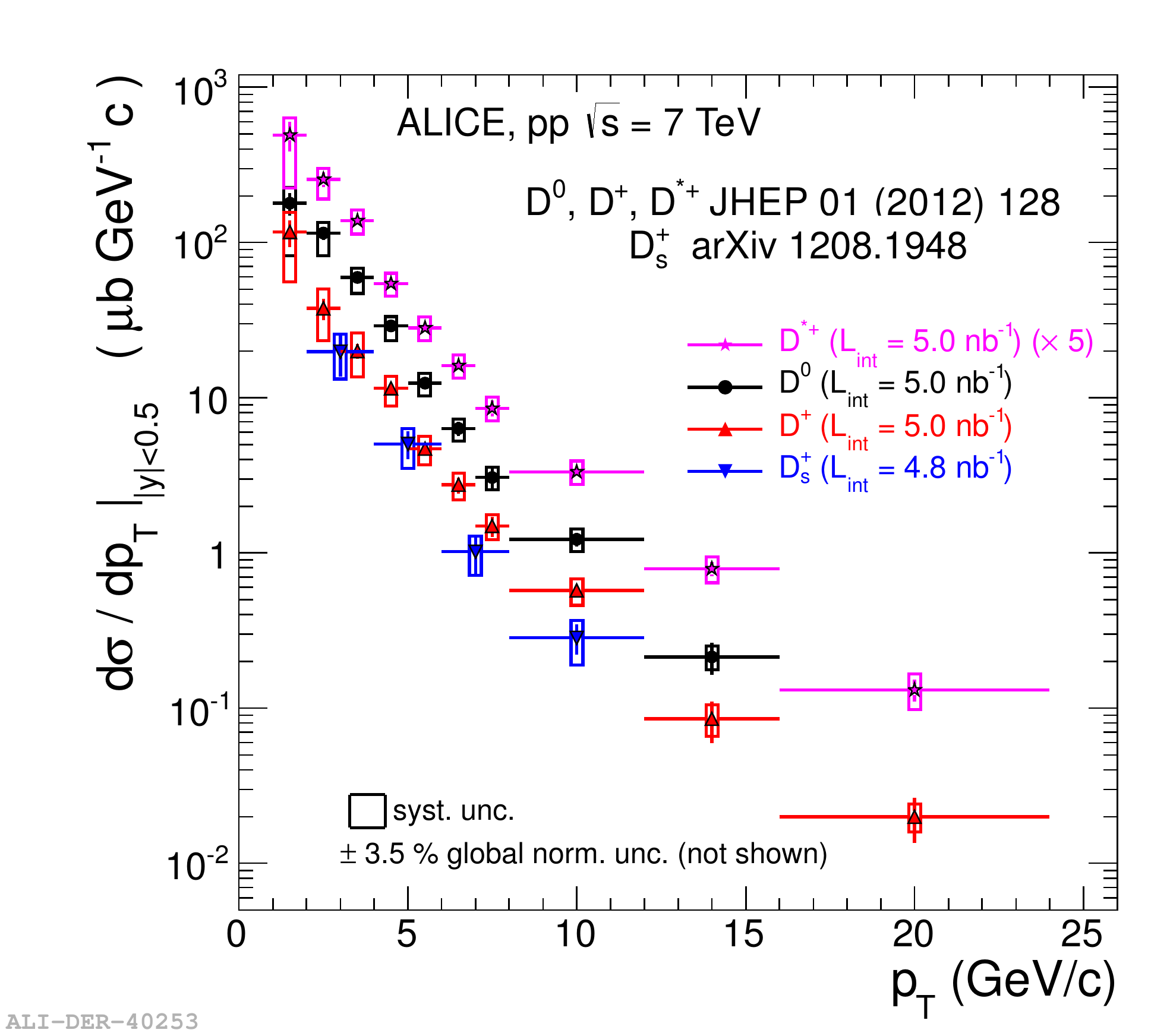}
\hspace{5mm}
\includegraphics[width=7.71cm,height=6.41cm]{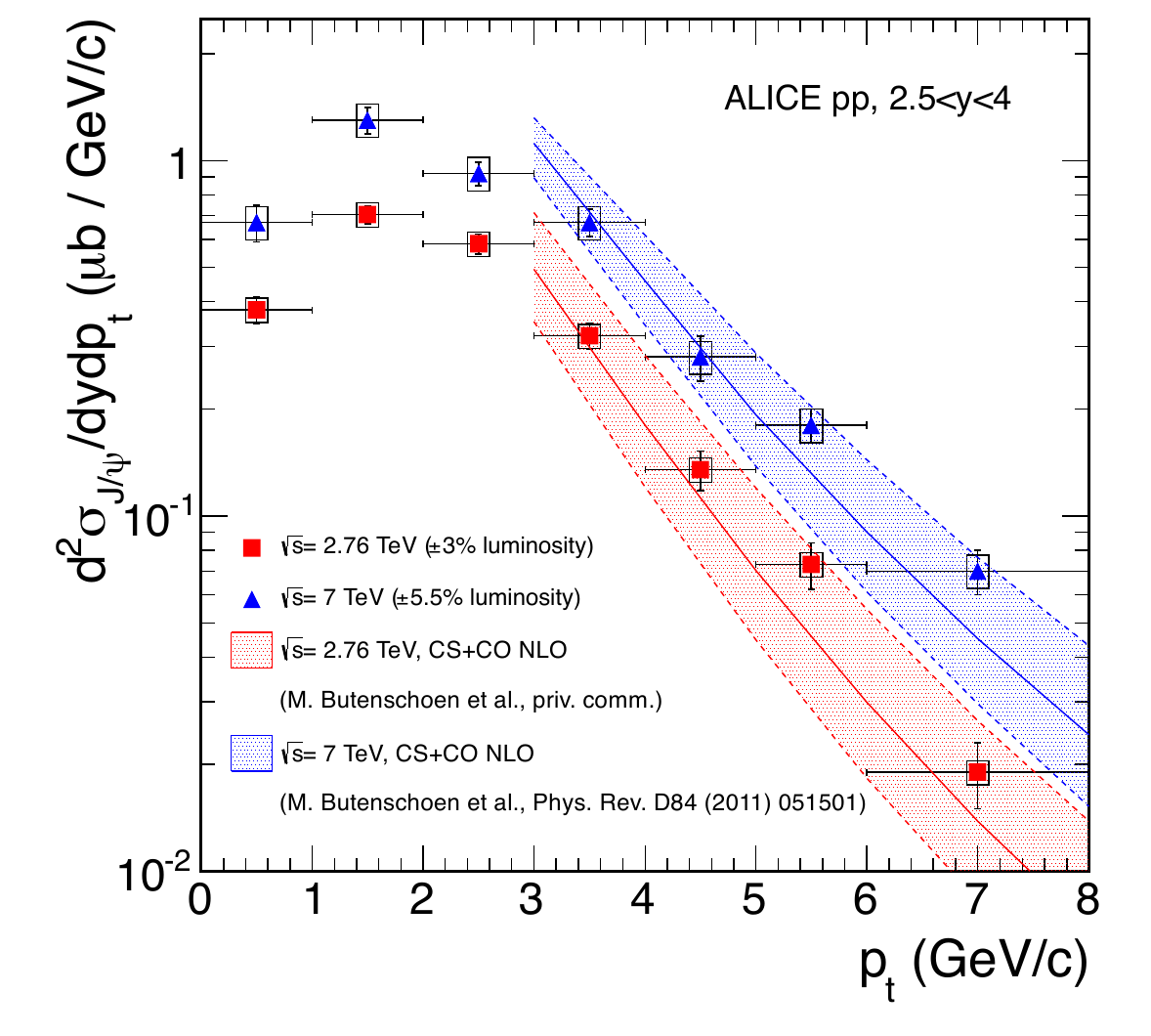}
\hspace{3mm}
\caption{(left) $p_{\rm T}$ -differential inclusive cross section for prompt D-mesons in pp at $\sqrt{s}$ = 7 TeV ~\cite{D7-key}. (right) Double differential J/$\psi$ production cross section at $\sqrt{s}$ = 7 and 2.76 TeV. The results are compared with a NLO NRQCD calculation performed in the region $p_{\rm T}$ $>$ 3 GeV/c ~\cite{Abelev:2012kr}.}
\label{fig1}
\end{center}
\end{figure}

             
The measurement of charmed D-mesons~\cite{D7-key,D276-key} (D$^{0}$, D$^{+}$, D$^{*+}$ and D$_{s}^{+}$) is performed via the topological reconstruction of the following decay channels: D$^{0}$$\rightarrow$K$^{-}$$\pi^{+}$, D$^{+}$$\rightarrow$K$^{-}$$\pi^{+}$$\pi^{+}$, D$^{*+}$$\rightarrow$D$^{0}$$\pi^{+}$, D$_{s}^{+}$$\rightarrow$$\phi$$\pi^{+}$, and their charge conjugates. Tracking is performed using the ITS and TPC detectors, and particle identification is made using the combined information from TPC and TOF. Finally, the reconstruction of the secondary vertices, required for the D-meson signal extraction from the large combinatorial background, is made using the ITS detector.         
The $p_{\rm T}$-differential inclusive cross sections of the prompt (B feed-down subtracted) charmed mesons D$^{0}$, D$^{+}$ and D$^{*+}$ in the rapidity range $\mid$y$\mid$~$<$~0.5\footnote{The analysis of open heavy flavours is performed in rapidity regions smaller than the one coverer by the central barrel in order to have an uniform acceptance.} were measured in pp collisions at $\sqrt{s}$ = 7 and 2.76 TeV. At $\sqrt{s}$ =~7~TeV, the D-meson analysis, performed on a data sample with an integrated luminosity $L_{int}$ = 5 $\rm nb^{-1}$, covers the $p_{\rm T}$ range 1 $<$ $p_{\rm T}$ $<$ 24 (16 for D$^{0}$) GeV/c. While at $\sqrt{s}$ = 2.76 TeV, the study is performed on a data sample with $L_{int}$~=~1.1~$\rm nb^{-1}$ over the $p_{\rm T}$ range 2~$<$~$p_{\rm T}$~$<$~12~GeV/c (1 $<$ $p_{\rm T}$ $<$ 12 GeV/c for the D$^{0}$). D$_{s}^{+}$ mesons have been measured at $\sqrt{s}$ = 7 TeV in the $p_{\rm T}$ range 2 $<$ $p_{\rm T}$ $<$ 12 GeV/c at mid-rapidity ($\mid$y$\mid$ $<$ 0.5) for a data sample with $L_{int}$ = 4.8 $\rm nb^{-1}$\cite{DS-key}. In Figure~\ref{fig1} (left) the $p_{\rm T}$ -differential inclusive cross sections for prompt D-mesons in pp collisions at $\sqrt{s}$~=~7~TeV are shown. The fraction of directly produced D-mesons, not fed down from B meson decays, was evaluated using the B production cross section from the FONLL pQCD calculations~\cite{FONLL1,FONLL2,FONLL3}. The fraction is estimated to be in a range between 80$\%$ and 90$\%$ depending on the $p_{\rm T}$ interval. The measured D-meson production differential cross sections are compared to two theoretical predictions, namely FONLL and GM-VFNS~\cite{GM1,GM2}. Both calculations are compatible with the measurements within the uncertainties. Finally, the total charm production cross sections at $\sqrt{s}$~=~7 and 2.76~TeV are obtained by extrapolating the data to the full phase space using the FONLL calculations. See ~\cite{D276-key} for detailed discussions on the total charm cross sections.

The measurement of electrons from semileptonic heavy flavour hadron decays has been performed in ALICE~\cite{Collaboration:2012rt} at mid-rapidity ($\mid$y$\mid$ $<$ 0.5) in pp collisions at $\sqrt{s}$ = 7 TeV in the transverse momentum range 0.5 $<$ $p_{\rm T}$ $<$ 8 GeV/c. The first step in the analysis is the electron identification which is provided by means of several detectors (TOF-TRD-TPC-EMCal). Then the electron background from sources other than semileptonic heavy flavour hadron decays is calculated using a Monte Carlo cocktail approach and subtracted from the inclusive electron spectra. Within the experimental and theoretical uncertainties FONLL is in agreement with the ALICE measurements. The production cross section of electrons from semileptonic decays of beauty hadrons has been also measured with the ALICE experiment at mid-rapidity ($\mid$y$\mid$ $<$ 0.8) in the transverse momentum range 1 $<$ $p_{\rm T}$ $<$ 8 GeV/c. Electrons from beauty-hadron decays were selected based on the displacement of the decay vertex from the pp collision (primary) vertex. A pQCD FONLL calculation has been compared to data and agrees with the measurement within uncertainties.    


In the forward rapidity region (2.5 $<$ y $<$ 4)\footnote{The ALICE muon spectrometer covers a negative $\eta$ range and, consequently, a negative y range. However, since in pp collisions the physics is symmetric with respect to y = 0, the negative sign when quoting the rapidity values has been dropped.}, the production of muons from heavy flavour hadron decays in pp collisions at $\sqrt{s}$ = 7 and 2.76 TeV was measured~\cite{Abelev:2012pi,Abelev:2012qh}. The analysis is carried out on a data sample with $L_{int}$ = 16.5 and 19 $\rm nb^{-1}$, respectively. The muon identification is performed by means of the muon trigger system, behind the iron wall, which allows to correctly select the muon candidates (with 99.9$\%$ purity). The subtraction of the background component, whose main sources are decays of primary pions and kaons, is based on Monte Carlo simulation (PYTHIA and PHOJET). The $p_{\rm T}$- and y-differential cross~sections of muons from heavy flavour decays (in the $p_{\rm T}$ range 2 $<$ $p_{\rm T}$ $<$ 12 GeV/c at $\sqrt{s}$ = 7 TeV and 2 $<$ $p_{\rm T}$ $<$ 10 GeV/c at $\sqrt{s}$ = 2.76 TeV) have been compared to FONLL pQCD predictions. The ALICE results are well reproduced by the calculations within experimental and theoretical uncertainties.

\section{Results on J/$\psi$ production}
\label{12}
In ALICE, the J/$\psi$ production can be studied in the $e^{+}$$e^{-}$ decay channel in the central rapidity region ($\mid$y$\mid$ $<$ 0.9), while the $\mu^{+}$$\mu^{-}$ decay channel is accessible in the forward rapidity region (2.5 $<$ y $<$ 4). The analysis of the J/$\psi$ production in pp collisions at $\sqrt{s}$ = 7 TeV (2.76 TeV), is performed, in the dimuon channel, on a data sample with $L_{int}$ = 15.6 $\rm nb^{-1}$ (19.9 $\rm nb^{-1}$), and in the dielectron channel with $L_{int}$ = 5.6 $\rm nb^{-1}$ (1.1 $\rm nb^{-1}$).
 

\begin{figure}[htbp]
\begin{center}

\includegraphics[width=7.cm,height=6.5cm]{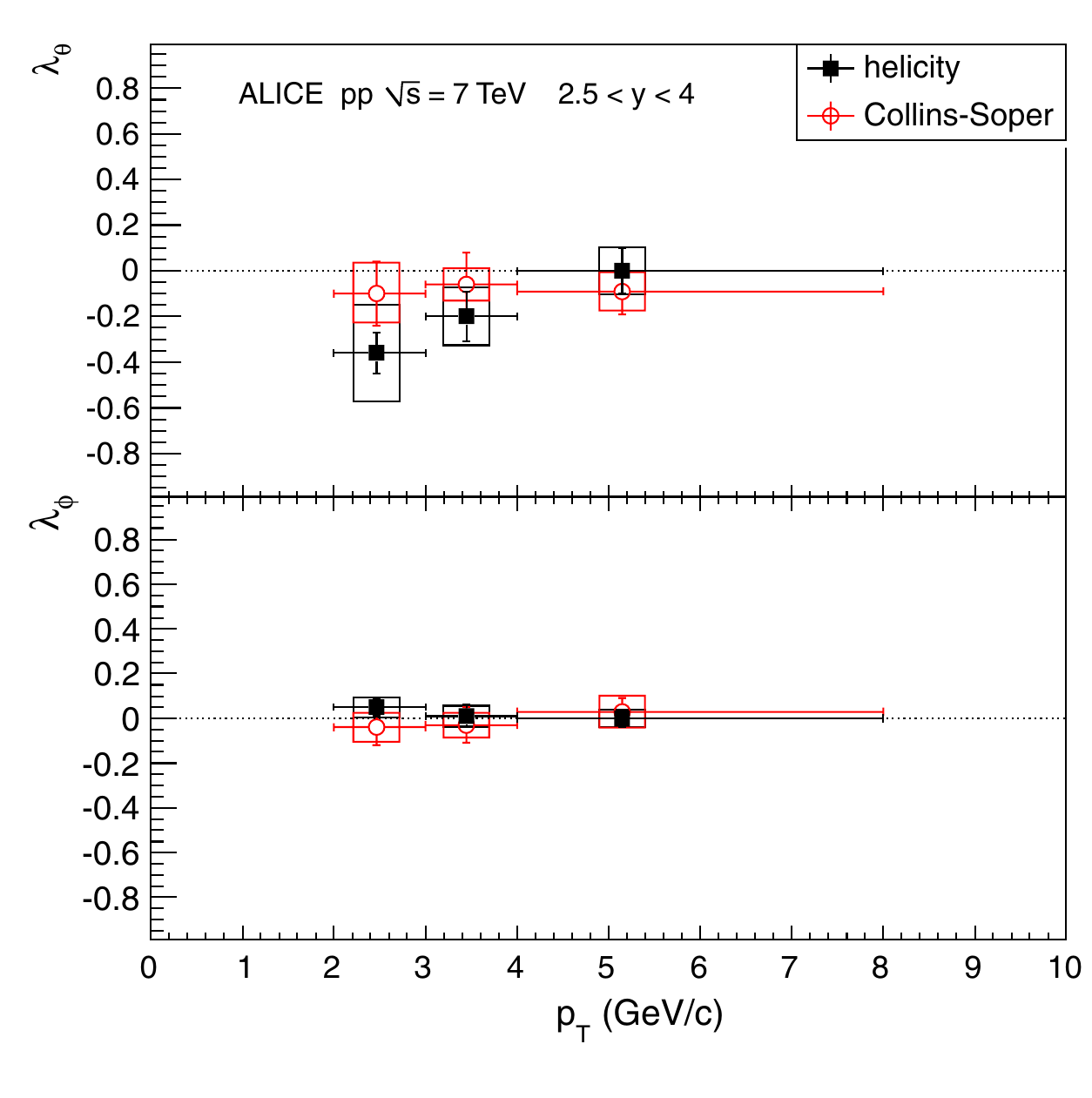}
\hspace{5mm}
\includegraphics[width=7.1cm,height=6.49cm]{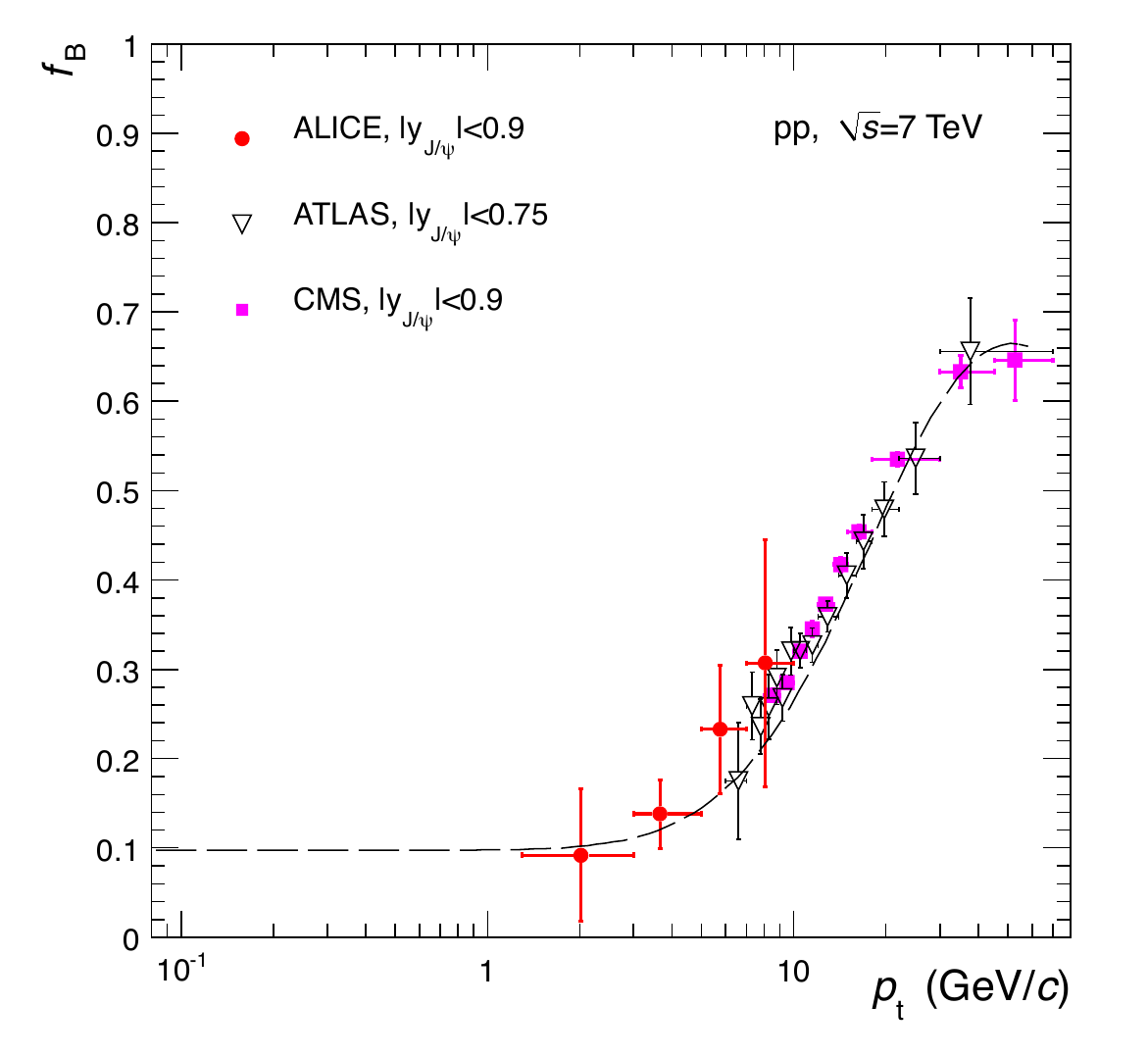}
\hspace{3mm}

\caption{(left) $\lambda_{\theta}$ and $\lambda_{\phi}$ parameter as a function of $p_{\rm T}$ for inclusive J/$\psi$, measured in the HE (closed squares) and CS (open circles) reference frames~\cite{pola}.   (right) The fraction of J/$\psi$ from the decay of b-hadrons as a function of $p_{\rm T}$ of J/$\psi$ compared with other LHC results~\cite{Abelev:2012rz}.}
\label{fig2}
\end{center}
\end{figure}

The inclusive d$\sigma_{J/\psi}$/dy cross sections in the two rapidity ranges, for the two energies, were measured~\cite{Aamodt:2011gj,Abelev:2012kr}. In the forward-y range, the $p_{\rm T}$-differential cross section was measured in the transverse momentum range 0 $<$ $p_{\rm T}$ $<$ 8 GeV/c. The d$^{2}\sigma_{J/\psi}$/dyd$p_{\rm T}$ for the two energies is shown in Figure~\ref{fig1} (right) where the two distributions are compared, in the region $p_{\rm T}$ $>$ 3 GeV/c, with a NLO NRQCD calculation which reproduces well the experimental data.\par


The study of the J/$\psi$$\rightarrow$$\mu^{+}$$\mu^{-}$ events in pp collisions at $\sqrt{s}$ = 7 TeV has also allowed to measure the polar ($\theta$) and azimuthal ($\phi$) angle distribution of the decay muons. From this study, performed in the region 2.5~$<$~y~$<$~4 and 2~$<$~$p_{\rm T}$~$<$~8~GeV/c, the J/$\psi$ polarization parameters $\lambda_{\theta}$ and $\lambda_{\phi}$ can be obtained in two different reference frames: the helicity (HE) and the Collins-Soper (CS) frame (see~\cite{pola} for details). Figure~\ref{fig2} (left) shows the results for inclusive J/$\psi$ production. The two polarization parameters  are consistent with zero, within the uncertainties, in both frames.\par

The J/$\psi$ production at $\sqrt{s}$ = 7 TeV has also been studied as a function of the relative charged particle pseudo-rapidity density, measured in the ITS ($\mid$y$\mid$ $<$ 1.6), both in the dielectron and in the dimuon channel~\cite{Abelev:2012rz}. The highest charged particle multiplicity density that we reach in this analysis is similar to the one measured in semi-peripheral Cu-Cu collisions at $\sqrt{s}_{NN}$ = 200 GeV (RHIC). Therefore, similarities between pp collisions at LHC and Cu-Cu collisions at RHIC might be found. ALICE measurements indicate a substantial increase (linear) of the relative (with respect to MB pp yield) J/$\psi$ yield as a function of the relative multiplicity. This is a non-trivial result since some theoretical calculations (PYTHIA 6.4) predicts the opposite effect.\par
Finally, we present the ALICE measurement of the fraction $f_{B}$ of the J/$\psi$ coming from the decay of long-lived beauty hadrons in pp collisions at $\sqrt{s}$ = 7 TeV ~\cite{Abelev:2012gx}. The kinematic region that can be explored, down to low $p_{\rm T}$ ($p_{\rm T} $$>$~1.3~GeV/c) at mid-rapidity ($\mid$y$\mid$ $<$ 0.9), is unique at the LHC. This analysis is feasible in the central barrel thanks to the good vertex position resolution which allows to distinguish prompt J/$\psi$ (decaying at the primary vertex) from non-prompt J/$\psi$ (decaying at a secondary vertex displaced respect to the primary vertex). Figure~\ref{fig2} (right) shows the fraction of J/$\psi$ from the decay of b-hadrons as a function of the J/$\psi$ $p_{\rm T}$ compared with other LHC results in pp collisions at $\sqrt{s}$ = 7 TeV. As can be seen, ALICE provides the data points at low $p_{\rm T}$ not accessible by other experiments.






\bibliographystyle{elsarticle-num}
\bibliography{BiblioProce}









\end{document}